%% file: as12n.TEX
\documentstyle[twoside,fleqn,espcrc2]{article}

\def\beq{\begin{equation}}
\def\eeq{\end{equation}}
\def\beqa{\begin{eqnarray}}
\def\eeqa{\end{eqnarray}}
\def\lla{\left\langle}
\def\rra{\right\rangle}
\def\lsim{\mathrel{\raise.3ex\hbox{$<$\kern-.75em\lower1ex\hbox{$\sim$}}} }
\def\gsim{\mathrel{\raise.3ex\hbox{$>$\kern-.75em\lower1ex\hbox{$\sim$}}} }
\newcommand{\za}{\alpha}
\newcommand{\zb}{\beta}

\newcommand{\AmS}{{\protect\the\textfont2
  A\kern-.1667em\lower.5ex\hbox{M}\kern-.125emS}}

\title{The Generic Supersymmetric Standard Model \\
\hspace*{.2in}as the Complete Theory of Supersymmetry without R-parity}

\author{Otto C. W. Kong
\address{Institute of Physics, Academia Sinica, Nankang, Taipei TAIWAN 11529}
}
       
\begin{document}

\thispagestyle{empty}

\onecolumn

\begin{flushright}
IPAS-HEP-k012\\
Feb 2001\\
ed. Apr 2001\\
\end{flushright}

\vspace*{.5in}

\begin{center}
{\bf  The Generic Supersymmetric Standard Model \\
\hspace*{.2in}as the Complete Theory of Supersymmetry without R-parity$^\star$ }\\
\vspace*{.5in}
{\bf  Otto C.W. Kong}\\[.05in]
{\it Institute of Physics, Academia Sinica, Nankang, Taipei, TAIWAN 11529}

\vspace*{.8in}
{Abstract}\\
\end{center}

The generic supersymmetric standard model is a model built from a supersymmetrized
standard model field spectrum the gauge symmetries only. The popular minimal 
supersymmetric standard model differs from the generic version in having
R-parity imposed by hand. We review an efficient formulation of the model
in which all the admissible R-parity violating terms are incorporated without 
bias. The model gives many new interesting R-parity violating phenomenological
features only started to be studied recently. Some of our recent results 
will be discussed, including newly identified 1-loop contributions
to neutrino masses and electric dipole moments of neutron and
electron. This is related to the largely overlooked R-parity violating 
contributions to squark and slepton mixings, which we also present in detail.

\vfill
\noindent --------------- \\
$^\star$ Talk the Thirty Years of SUSY Symposium and Workshop 
(Oct 13 - 27, 2000), Theoretical Physics Institute,
University of Minnesota, USA\\
 --- submission for the proceedings.  
 
\clearpage
\addtocounter{page}{-1}
\input{as12}

\end{document}

%% file: as12.tex
\begin{abstract}
The generic supersymmetric standard model is a model built from a supersymmetrized
standard model field spectrum the gauge symmetries only. The popular minimal 
supersymmetric standard model differs from the generic version in having
R-parity imposed by hand. We review an efficient formulation of the model
in which all the admissible R-parity violating terms are incorporated without 
bias. The model gives many new interesting R-parity violating phenomenological
features only started to be studied recently. Some of our recent results 
will be discussed, including newly identified 1-loop contributions
to neutrino masses and electric dipole moments of neutron and
electron. This is related to the largely overlooked R-parity violating 
contributions to squark and slepton mixings, which we also present in detail.
\end{abstract}

\maketitle

\section{INTRODUCTION}
From the early history of supersymmetry (SUSY), there had been thinkings about
its usage in the obviously non-supersymmetric low-energy phenomenology. One of the 
first idea was the identification of the neutrino as a goldstino, {\it i.e.} 
the Goldstone mode from (global) SUSY breaking\cite{AV}. Nowadays, the 
question : ``Is the masslessness of the neutrino a result of SUSY (breaking)?"
is obvious an uninteresting one. Nevertheless, {neutrinos} and {SUSY} just
may have everything to do with one another; afterall, {\it nonzero} masses of 
neutrinos may be a result of SUSY. The latter is related to the notion of
R-parity violation --- the topic discussed here.

The notion of R-parity came about also  early in the history of SUSY\cite{Fy2}.
In those days, baryon and lepton number symmetries might look even better
than the standard model (SM) itself. R-parity, being basically baryon and lepton 
number symmetries of a supersymmetric SM, seemed quite natural. However, 
global symmetries are understood to be less than sacred. The basic
theoretical building block of the SM is the field spectrum and the gauge
symmetries, while baryon and lepton number symmetries come out only as
an accident. In fact, there are now strong evidence of nonzero neutrino 
masses\cite{osc}. Moreover, such evidence is our only definite experimental 
indication of existence particle physics beyond the SM. On the contrary, 
SUSY itself still awaits discovery --- that is, if it has anything to do with
nature at all. Most, if not all, 
neutrino mass models actually violate lepton number symmetry. Of course if one
simply adds (softly broken) SUSY to the basic building blocks of the SM, the
(generic) supersymmetric SM thus obtained admits violations of
baryon and lepton number symmetries and nonzero neutrino masses. There is the
acute problem of superparticle mediated proton decay. But even in the 
consideration of the issue, R-parity certainly overkills. It is not the only 
candidate for the job, nor is it the most effective\cite{pd}.
It is the most restrictive though, in terms of what terms are admitted in the 
renormalizable Lagrangian and otherwise. May be the only
advantage of R-parity is to provide a much simpler model for phenomenological
analyses --- the minimal supersymmetric SM (MSSM). However, the generic 
supersymmetric SM is, at least conceptually, the simplest model with SUSY and
neutrino masses. Here, the latter can in fact be considered as a consequence of
supersymmetrizing the SM. Hence, we take a simple phenomenological perspective 
here, taking the generic supersymmetric SM and study the experimental constraints 
on the various couplings with {\it a priori} bias. From the theoretical point of
view, some sort of baryon number, in relation to proton decay, is expected to 
be protected by symmetry, while lepton numbers are likely to be violated.

\section{THE GENERIC SUPERSYMMETRIC STANDARD MODEL}
The most general renormalizable superpotential for the generic supersymmetric
SM (without R-parity) can be written  as
\small\beqa
W \!\!\!\! &=& \!\!\!\!\varepsilon_{ab}\Big[ \mu_{\alpha}  \hat{H}_u^a \hat{L}_{\alpha}^b 
+ h_{ik}^u \hat{Q}_i^a   \hat{H}_{u}^b \hat{U}_k^{\scriptscriptstyle C}
+ \lambda_{\alpha jk}^{\!\prime}  \hat{L}_{\alpha}^a \hat{Q}_j^b
\hat{D}_k^{\scriptscriptstyle C} 
\nonumber \\
\!\!\!\!&& \!\! +\;
\frac{1}{2}\, \lambda_{\alpha \beta k}  \hat{L}_{\alpha}^a  
 \hat{L}_{\beta}^b \hat{E}_k^{\scriptscriptstyle C} \Big] + 
\frac{1}{2}\, \lambda_{ijk}^{\!\prime\prime}  
\hat{U}_i^{\scriptscriptstyle C} \hat{D}_j^{\scriptscriptstyle C}  
\hat{D}_k^{\scriptscriptstyle C}   ,
\eeqa\normalsize
where  $(a,b)$ are $SU(2)$ indices, $(i,j,k)$ are the usual family (flavor) 
indices, and $(\za, \zb)$ are extended flavor index going from $0$ to $3$.
In the limit where $\lambda_{ijk}, \lambda^{\!\prime}_{ijk},  
\lambda^{\!\prime\prime}_{ijk}$ and $\mu_{i}$  all vanish, 
one recovers the expression for the R-parity preserving case, 
with $\hat{L}_{0}$ identified as $\hat{H}_d$. Without R-parity imposed,
the latter is not {\it a priori} distinguishable from the $\hat{L}_{i}$'s.
Note that $\lambda$ is antisymmetric in the first two indices, as
required by  the $SU(2)$  product rules, as shown explicitly here with 
$\varepsilon_{\scriptscriptstyle 12} =-\varepsilon_{\scriptscriptstyle 21}=1$.
Similarly, $\lambda^{\!\prime\prime}$ is antisymmetric in the last two 
indices, from $SU(3)_{\scriptscriptstyle C}$. 

R-parity is exactly an {\it ad hoc} symmetry put in to make $\hat{L}_{0}$,
stand out from the other $\hat{L}_i$'s as the candidate for  $\hat{H}_d$.
It is defined in terms of baryon number, lepton number, and spin as, 
explicitly, ${\mathcal R} = (-1)^{3B+L+2S}$. The consequence is that 
the accidental symmetries of baryon number and lepton number in the SM 
are preserved, at the expense of making particles and superparticles having 
a categorically different quantum number, R-parity. As mentioned above, R-parity
hence kills the dangerous proton decay but also forbides neutrino masses
within the model.

There are certainly no lack of studies on various ``R-parity violating models" 
in the literature. However, such models typically involve strong assumptions
on the form of R-parity violation. In most cases, no clear statement on
what motivates the assumptions taken is explicitly given. In fact, there are 
quite some confusing, or even plainly wrong, statements on the issues concerned.
It is important to distinguish among the different RPV ``theories", and, 
especially, between such a theory and the unique generic supersymmetric 
SM\cite{as8}. The latter {\it is} the {\it complete} theory of SUSY without 
R-parity, one which admits all the RPV terms without {\it a priori} bias. 

R-parity violating (RPV) parameters come in various froms. These include
the more popular trilinear ($\lambda_{ijk}$, $\lambda_{ijk}^{\prime}$,
and	$\lambda_{ijk}^{\prime\prime}$) and bilinear ($\mu_i$)
couplings in the superpotential, as well as  soft SUSY breaking
parameters of the trilinear, bilinear, and soft mass (mixing) types.
From a phenomenological point of view, there is the related notion
of (RPV) ``sneutrino VEV's". In order not to miss any plausible RPV
phenomenological features, it is important that all of the RPV
parameters be taken into consideration. For example, they all have a role to play
in neutrino mass generations\cite{as9}. The soft SUSY breaking part 
of the Lagrangian is more interesting, if only for the fact that  many
of its interesting details have been overlooked in the literature.
However, we will postpone the discussion for the moment, to the latter part of 
the article.

\subsection{Supersymmetrizing the Standard Model}
Let us review here the supersymmetrization of 
the SM. In the matter field sector, all fermions and scalars have to
be promoted to chiral superfields containing both parts. It is 
straight forward for the quark doublets and singlets, and also for the leptonic
singlet. The leptonic doublets, however, have the same quantum number as
the Higgs doublet that couples to the down-sector quarks. Nevertheless, one
cannot simply get the Higgs, $H_d$, from the scalar partners of the leptonic 
doublets, $L$'s. Holomorphicity of the superpotential requires a separate superfield 
to contribute the Higgs coupling to the up-sector quarks. This $\hat{H}_u$ 
superfield then contributes a fermionic doublet, the Higgsino, with non-trivial
gauge anomaly. To cancel the latter, an extra fermionic doublet with
the quantum number of $H_d$ or $L$ is needed. So, the result is that we need
four superfields with that quantum number. As they are {\it a priori} 
indistinguishable, we label them by $\hat{L}_{\alpha}$ ($\za=0$ to $3$). With
the superfield content and the SM gauge symmetries, we have the superpotential
given above.

\subsection{The Single-VEV Parametrization}
After the supersymmetrization, however, some of the superfields lose the exact
identities they have in relation to the physical particles. The latter has to
be mass eigenstates, which have to be worked out from the Lagrangian of the model.
Assuming electroweak symmetry breaking, we have now five (color-singlet) charged
fermions, for example. There are also 1+4 VEV's admitted, together with a SUSY
breaking gaugino mass. If one writes down naively the (tree-level) mass matrix,
the result is extremely complicated (see \cite{ru10} for an explicit illsutration),
with all the $\mu_{\scriptscriptstyle \za}$ and $\lambda_{\za\zb k}$ couplings
involved, from which the only definite experimental data are the three physical
lepton masses as the light eigenvalues, and the overall magnitude of
the electroweak symmetry breaking VEV's. The task of analyzing the model seems
to be formidable.

Doing phenomenological studies without specifying a choice 
of flavor bases is, however, ambiguous. It is like doing SM quark physics with 18
complex Yukawa couplings, instead of the 10 real physical parameters.
As far as the SM itself is concerned, the extra 26 real parameters
are simply redundant, and attempts to relate the full 36 parameters to experimental 
data will be futile. In the case at hand, the choice of an optimal
parametrization mainly concerns the 4 $\hat{L}_\alpha$ flavors. We use
here the single-VEV parametrization\cite{ru} (SVP), in which flavor bases 
are chosen such that : 
1/ among the $\hat{L}_\alpha$'s, only  $\hat{L}_0$, bears a VEV,
{\it i.e.} {\small $\langle \hat{L}_i \rangle \equiv 0$};
2/  {\small $h^{e}_{jk} (\equiv \lambda_{0jk}) 
=\frac{\sqrt{2}}{v_{\scriptscriptstyle 0}} \,{\rm diag}
\{m_{\scriptscriptstyle 1},
m_{\scriptscriptstyle 2},m_{\scriptscriptstyle 3}\}$};
3/ {\small $h^{d}_{jk} (\equiv \lambda^{\!\prime}_{0jk} =-\lambda_{j0k}) 
= \frac{\sqrt{2}}{v_{\scriptscriptstyle 0}}\,{\rm diag}\{m_d,m_s,m_b\}$}; 
4/ {\small $h^{u}_{ik}=\frac{\sqrt{2}}{v_{\scriptscriptstyle u}}
V_{\mbox{\tiny CKM}}^{\!\scriptscriptstyle T} \,{\rm diag}\{m_u,m_c,m_t\}$}, where 
${v_{\scriptscriptstyle 0}} \equiv  \sqrt{2}\,\langle \hat{L}_0 \rangle$
and ${v_{\scriptscriptstyle u} } \equiv \sqrt{2}\,
\langle \hat{H}_{u} \rangle$. Thus, the parametrization singles out the
$\hat{L}_0$ superfield as the one containing the Higgs. As a result,
it gives the complete RPV effects on the {tree-level mass matrices} 
of all the states (scalars and fermions) the simplest structure.
The latter is a strong technical advantage.

\section{The (Color-singlet) Fermions}
The SVP gives quark mass matrices exactly in the SM form. For the masses
of the color-singlet fermions, all the RPV effects are paramatrized by the
$\mu_i$'s only. For example, the five charged fermions (gaugino + higgsino +  
3 charged leptons), we have
\small\beq \label{mc}
{\mathcal{M}_{\scriptscriptstyle C}} =
 \left(
{\begin{array}{ccccc}
{M_{\scriptscriptstyle 2}} &  
\frac{g_{\scriptscriptstyle 2}{v}_{\scriptscriptstyle 0}}{\sqrt 2}  
& 0 & 0 & 0 \\
 \frac{g_{\scriptscriptstyle 2}{v}_{\scriptscriptstyle u}}{\sqrt 2} & 
 {{ \mu}_{\scriptscriptstyle 0}} & {{ \mu}_{\scriptscriptstyle 1}} &
{{ \mu}_{\scriptscriptstyle 2}}  & {{ \mu}_{\scriptscriptstyle 3}} \\
0 &  0 & {{m}_{\scriptscriptstyle 1}} & 0 & 0 \\
0 & 0 & 0 & {{m}_{\scriptscriptstyle 2}} & 0 \\
0 & 0 & 0 & 0 & {{m}_{\scriptscriptstyle 3}}
\end{array}}
\right)  \; .
\eeq\normalsize
Moreover each $\mu_i$ parameter here characterizes directly the RPV effect
on the corresponding charged lepton  ($\ell_i = e$, $\mu$, and $\tau$) \cite{ru}.
Hence, in the limit of small $\mu_i$'s (relative to $M_{\scriptscriptstyle 2}$ and
${{\mu}_{\scriptscriptstyle 0}}$), the superfields $\hat{L}_i$'s and 
$\hat{E}_k^{\scriptscriptstyle C}$'s have small deviations from the $\ell_i$
superfields, with $m_i$'s being roughly the physical masses. In general,
for any set of other parameter inputs, the ${m}_i$'s can still be determined,
through a numerical procedure, to guarantee that the correct mass
eigenvalues of  $m_e$, $m_\mu$, and $m_\tau$  are obtained --- an issue
first addressed and solved in Ref.\cite{ru}.

Under the SVP,  neutral fermion (neutralino-neutrino) mass matrix  
has  also RPV contributions from the three $\mu_i$'s only. The mass matrix
can be written in the $3+4$ block form
\begin{equation} \label{mn}
{\cal{M}_{\scriptscriptstyle N}} =
\left(
\begin{array}{cc}
\cal{M} & \xi^{\scriptscriptstyle T} \\
\xi & m_\nu^{\scriptscriptstyle 0}
\end{array}
 \right) \; ,
\eeq
where, at the tree-level,
\beqa
\cal{M}
\!\!\!\! &=& \!\!\!\! 
\left(
\begin{array}{cccc}
{{M}_{\scriptscriptstyle 1}} & 0 &  \frac {{g}_{1}{v}_{u}}{2}
 &  -\frac{{g}_{1}{v}_{d}}{2}  \\
0 & {{M}_{\scriptscriptstyle 2}} &  -\frac{{g}_{2}{v}_u}{2} & 
\frac{{g}_{2}{v}_{d}}{2} \\
 \frac {{g}_{1}{v}_{u}}{2} &   -\frac{{g}_{2}{v}_u}{2} 
& 0 &  - {{\mu}_{\scriptscriptstyle 0}}\\
 -\frac{{g}_{1}{v}_{d}}{2} & \frac{g_{2}{v}_d}{2}
 &  - {{ \mu}_{0}} & 0
\end{array} \right)  \; ,
\nonumber \\
\xi 
\!\!\!\! &=& \!\!\!\! 
\left(
\begin{array}{cccc}
0 & 0 &  - {{ \mu}_{\scriptscriptstyle 1}} & 0 \\
0 & 0 &  - {{ \mu}_{\scriptscriptstyle 2}} & 0 \\
0 & 0 &  - {{ \mu}_{\scriptscriptstyle 3}} & 0 
\end{array}
\right)  \; ,
\nonumber \\
m_\nu^{\scriptscriptstyle 0}
\!\!\!\! &=& \!\!\!\! 
 0_{\scriptscriptstyle 3 \times 3} \; .
\eeqa
For small $\mu_i$'s, it has a ``seesaw"  type structure,  with the effective
neutrino mass matrix given by
\beq
m_\nu =
- \xi {\cal{M}}^{-1}  \xi^{\scriptscriptstyle T}  \; .
\eeq
with one non-zero mass eigenvalue given as
\small \[
-  \frac {1}{2}
\frac{  {v}^{2} \cos^2\!\!\zb 
\left( {g}_{\scriptscriptstyle 2}^{2}\,M_{\scriptscriptstyle 1}
+ {g}_{\scriptscriptstyle 1}^{2}\,M_{\scriptscriptstyle 2} \right) 
\;\; \mu_{\scriptscriptstyle 5}^2}
{\mu_{\scriptscriptstyle 0} \left[ 2\,M_{\scriptscriptstyle 1}\,
M_{\scriptscriptstyle 2}
 \mu_{\scriptscriptstyle 0} -
 \left( {g}_{\scriptscriptstyle 2}^{2}\,M_{\scriptscriptstyle 1}
+{g}_{\scriptscriptstyle 1}^{2}\,M_{\scriptscriptstyle 2}\right) 
{v}^2 \sin\!{\zb}\cos\!{\zb} \right] } \; ;
\] \normalsize
where $\mu_{\scriptscriptstyle 5}^2 = \mu_{\scriptscriptstyle 1}^2
+ \mu_{\scriptscriptstyle 2}^2 + \mu_{\scriptscriptstyle 3}^2$
and the corresponding eigenstate is an admixture of the three basis 
neutrino states of $m_\nu$ here exactly in proportion 
$\frac{\mu_i}{\mu_{\scriptscriptstyle 5}}$.
Note that at the limit of small $\mu_i$'s, the three neutrino states  correspond to $\nu_e$, $\nu_\mu$, and $\nu_\tau$.

\section{Some Phenomenlogical Implications}
Taking the fermion mass matrices above and analyzing the resulted $Z^0$- and $W^\pm$-
couplings of the physical states, an interesting list of tree-level
RPV phenomenology from the gauge interactions can be exploited to get some
constraints on the $\mu_i$ parameters. The topic is studied in detail in 
Ref.\cite{ru}, which we summarized here in Table~1.
\input{as12t}

As for neutrino masses, apart from the tree-level contribution given above, there
are the well-studied 1-loop contributions from the $\lambda^{\!\prime}$- or
$\lambda$- couplings, with interesting implications on the flavor structure
of the classes of parameters\cite{ok}. There are also contributions involving
a bilinear together with a trilinear parameter, as first pointed out in the 
study\cite{as1} of a SUSY version of the Zee model\cite{zee}. Such kind of
contributions are closely related to RPV contributions to scalar mixings. The topic
is much overlooked till our recent analysis\cite{as5}.

The most interesting RPV contributions to scalar masses involve both the
bilinear and trilinear parameters, coming into the $LR$-mixings part of the
mass matrices. It is then very easy to see that they give rise to RPV 
contributions to electric dipole moments (EDM's)\cite{as4,cch1,as6}, 
as well the important flavor changing processes such as
$b\to s \, \gamma$\cite{bs} and $\mu \to e\, \gamma$\cite{as7}. In the
discussion below, we will illustrate some of these interesting recent results.

\section{SOFT TERMS AND SQUARKS}
\subsection{The Soft SUSY Breaking Terms}
The soft SUSY breaking part of the Lagrangian can be written as 
\footnotesize\beqa
&& \!\!\!\!\!\!\!\! V_{\rm soft}
= \tilde{Q}^\dagger \tilde{m}_{\!\scriptscriptstyle {Q}}^2 \,\tilde{Q} 
+\tilde{U}^{\dagger} 
\tilde{m}_{\!\scriptscriptstyle {U}}^2 \, \tilde{U} 
+\tilde{D}^{\dagger} \tilde{m}_{\!\scriptscriptstyle {D}}^2 
\, \tilde{D} +
 \tilde{L}^\dagger \tilde{m}_{\!\scriptscriptstyle {L}}^2  \tilde{L}  
 \nonumber \\
&+&
 \tilde{E}^{\dagger} \tilde{m}_{\!\scriptscriptstyle {E}}^2 
\, \tilde{E}
+ \tilde{m}_{\!\scriptscriptstyle H_{\!\scriptscriptstyle u}}^2 \,
|H_{u}|^2 
+  \Big[ \,
\frac{M_{\!\scriptscriptstyle 1}}{2} \tilde{B}\tilde{B}
   + \frac{M_{\!\scriptscriptstyle 2}}{2} \tilde{W}\tilde{W}
 \nonumber \\
&+&
\frac{M_{\!\scriptscriptstyle 3}}{2} \tilde{g}\tilde{g}
+\epsilon_{\!\scriptscriptstyle ab} \Big( \,
  B_{\za} \,  H_{u}^a \tilde{L}_\za^b 
+ A^{\!\scriptscriptstyle U}_{ij} \, 
\tilde{Q}^a_i H_{u}^b \tilde{U}^{\scriptscriptstyle C}_j 
 \nonumber \\
&+&
 A^{\!\scriptscriptstyle D}_{ij} 
H_{d}^a \tilde{Q}^b_i \tilde{D}^{\scriptscriptstyle C}_j  
+ A^{\!\scriptscriptstyle E}_{ij} 
H_{d}^a \tilde{L}^b_i \tilde{E}^{\scriptscriptstyle C}_j 
+ A^{\!\scriptscriptstyle \lambda^\prime}_{ijk} 
\tilde{L}_i^a \tilde{Q}^b_j \tilde{D}^{\scriptscriptstyle C}_k
\nonumber \\
&+& 
\frac{1}{2}\, A^{\!\scriptscriptstyle \lambda}_{ijk} 
\tilde{L}_i^a \tilde{L}^b_j \tilde{E}^{\scriptscriptstyle C}_k  \Big)
+ \frac{1}{2}\, A^{\!\scriptscriptstyle \lambda^{\prime\prime}}_{ijk}
 \tilde{U}^{\scriptscriptstyle C}_i  \tilde{D}^{\scriptscriptstyle C}_j  
\tilde{D}^{\scriptscriptstyle C}_k  
  +  \mbox{\normalsize h.c.} \Big]
\label{soft}
\eeqa\normalsize
where we have separated the R-parity conserving $A$-terms from the 
RPV ones (recall $\hat{H}_{d} \equiv \hat{L}_0$). Note that 
$\tilde{L}^\dagger \tilde{m}_{\!\scriptscriptstyle \tilde{L}}^2  \tilde{L}$,
unlike the other soft mass terms, is given by a 
$4\times 4$ matrix. Explicitly, 
$\tilde{m}_{\!\scriptscriptstyle {L}_{00}}^2$ corresponds to 
$\tilde{m}_{\!\scriptscriptstyle H_{\!\scriptscriptstyle d}}^2$ 
of the MSSM case while 
$\tilde{m}_{\!\scriptscriptstyle {L}_{0k}}^2$'s give RPV mass mixings.
Going from here, it is straight forward to obtain
the squark and slepton masses. 

\subsection{Down Squark Mixings}
The SVP also simplifies much the otherwise extremely complicated expressions
for the mass-squared matrices of the scalar sectors. Firstly, we will look 
at the squarks sectors. The masses of up-squarks obviously have no RPV 
contribution. The down-squark sector, however, has interesting result. 
We have the mass-squared matrix as follows : 
\beq \label{MD}
{\cal M}_{\!\scriptscriptstyle {D}}^2 =
\left( \begin{array}{cc}
{\cal M}_{\!\scriptscriptstyle LL}^2 & {\cal M}_{\!\scriptscriptstyle RL}^{2\dag} \\
{\cal M}_{\!\scriptscriptstyle RL}^{2} & {\cal M}_{\!\scriptscriptstyle RR}^2
 \end{array} \right) \; ,
\eeq
where
\small \beqa
{\cal M}_{\!\scriptscriptstyle LL}^2 
\!\!\!\! &=& \!\!\!\!
\tilde{m}_{\!\scriptscriptstyle {Q}}^2 +
m_{\!\scriptscriptstyle D}^\dag m_{\!\scriptscriptstyle D}
+ M_{\!\scriptscriptstyle Z}^2\, \cos\!2 \beta 
\left[ -\frac{1}{2} + \frac{1}{3} \sin\!^2 \theta_{\!\scriptscriptstyle W}\right]  ,
\nonumber \\
{\cal M}_{\!\scriptscriptstyle RR}^2 
\!\!\!\! &=& \!\!\!\!
\tilde{m}_{\!\scriptscriptstyle {D}}^2 +
m_{\!\scriptscriptstyle D} m_{\!\scriptscriptstyle D}^\dag
+ M_{\!\scriptscriptstyle Z}^2\, \cos\!2\beta 
\left[  - \frac{1}{3} \sin\!^2 \theta_{\!\scriptscriptstyle W}\right]  ,
\nonumber \\
\eeqa 
{\normalsize and}
\beqa 
({\cal M}_{\!\scriptscriptstyle RL}^{2})^{\scriptscriptstyle T} 
\!\!\!\! &=& \!\!\!\!
A^{\!{\scriptscriptstyle D}} \frac{v_{\scriptscriptstyle 0}}{\sqrt{2}}
- (\, \mu_{\scriptscriptstyle \za}^*\lambda^{\!\prime}_{{\scriptscriptstyle \za}jk}\,)
\; \frac{v_{\scriptscriptstyle u}}{\sqrt{2}} \; 
\nonumber \\ \!\!\!\! &=& \!\!\!\! 
\left[ A_d -  \mu_{\scriptscriptstyle 0}^* \, \tan\!\beta \right]
\,m_{\!\scriptscriptstyle D}\;
+ \frac{\sqrt{2}\, M_{\!\scriptscriptstyle W} \cos\!\beta}
{g_{\scriptscriptstyle 2} } \,
\delta\! A^{\!{\scriptscriptstyle D}}
\nonumber \\
\!\!\!\! && 
- \frac{\sqrt{2}\, M_{\!\scriptscriptstyle W} \sin\!\beta}
{g_{\scriptscriptstyle 2} } \,
(\, \mu_i^*\lambda^{\!\prime}_{ijk}\, ) \; .
\label{RL}
\eeqa \normalsize
Here, $m_{\!\scriptscriptstyle D}$ is the down-quark mass matrix, 
which is diagonal under the parametrization adopted; 
$A_d$ is a constant (mass) parameter representing the 
``proportional" part of the $A$-term and the matrix 
$\delta\! A^{\!{\scriptscriptstyle D}}$ is the ``proportionality" violating 
part; $(\, \mu_i^*\lambda^{\!\prime}_{ijk}\, )$, and similarly 
$(\, \mu_{\scriptscriptstyle \za}^*\lambda^{\!\prime}_{{\scriptscriptstyle \za}jk}\,)$, 
denotes the $3\times 3$ matrix $(\;)_{jk}$ with elements listed. The   
$(\, \mu_{\scriptscriptstyle \za}^*\lambda^{\!\prime}_{{\scriptscriptstyle \za}jk}\,)$
term is the full $F$-term contribution, while the 
$(\, \mu_i^*\lambda^{\!\prime}_{ijk}\, )$ part separated out in the last expression
gives the RPV contributions. It is important to note
that the term contains flavor-changing ($j\ne k$) parts which,
unlike the $A$-terms ones, cannot be suppressed through a flavor-blind
SUSY breaking spectrum; even for the diagonal part, the phase cannot be suppressed
as that of the $A$-terms from gauge mediation. The novel issue here is that the
RPV contributions come from supersymmetric, rather than SUSY breaking parameters.

\section{CONTRIBUTIONS TO NEUTRON EDM}
\subsection{The Illustrative Gaugino Loop \\ \ \ Contributions} 
It is familiar in SUSY phenomenology that diagonal $LR$-scalar mixings giving 
rise to EDM's. For the $d$ quark EDM, we have then a direct contribution coming 
from a gaugino loop. The diagram looks the same as the MSSM gluino
and neutralino diagram with two gauge coupling vertices. The new RPV contributions 
here is a simple result of the RPV $LR$ squark mixings [{\it cf.} Eq.(\ref{RL})].
In Ref.\cite{as4}, we focused on the illustrative gluino loop contribution; while
the complete SUSY loop contributions to neutron EDM are analyzed in detail in 
Ref.\cite{as6}, with a comprehensive exact numerical study. Notice that though 
both the $u$ and $d$ quarks get EDM from gaugino loops in MSSM, only the $d$
quark has the RPV contribution. The $u$-squark sector simply has no RPV 
$LR$ mixings. 

Neglecting inter-family mixings among the squarks, we have the gluino 
diagram contribution in the often quoted expression
\beq \label{G02}
\left({d_{\scriptscriptstyle d} \over e}\right)_{\!\!\tilde{g}} =
-{2 \alpha_s \over 3 \pi} \;
{M_{\!\scriptscriptstyle \tilde{g}} \over M^2_{\!\scriptscriptstyle \tilde{d}} } \; 
{\cal Q}_{\tilde{d}}\; 
\mbox{Im}(\delta^{\scriptscriptstyle D}_{\!\scriptscriptstyle 1\!1}) \;
F \!\left( {M_{\!\scriptscriptstyle \tilde{g}}^2 \over M^2_{\!\scriptscriptstyle \tilde{d}}} \right) \; ,
\eeq
where $\delta_{\!\scriptscriptstyle 1\!1}^{\scriptscriptstyle D}$ is 
${\cal M}_{\!\scriptscriptstyle RL}^2 / M_{\!\scriptscriptstyle \tilde{d}}^2$ 
(with ${\cal M}_{\!\scriptscriptstyle RL}^2$ restricted to the $\tilde{d}$ family)
and
\beq \label{Fx}
F(x) = {1 \over (1-x)^3} \left[ { 1+ 5 x \over 2} +
{(2+x) x \ln x \over (1-x)} \right] \; .
\eeq
The expression above is, in fact, the same as that of the MSSM case,
except that $\delta_{\!\scriptscriptstyle 1\!1}^{\scriptscriptstyle D}$,
or equivalently ${\cal M}_{\!\scriptscriptstyle RL}^2$, has now an extra 
RPV part. From the general result given in Eq.(\ref{RL}), we have for 
the $\tilde{d}$ squark,
\small \beqa \label{delta-D}
\delta_{\!\scriptscriptstyle 1\!1}^{\scriptscriptstyle D} 
 M^2_{\!\scriptscriptstyle \tilde{d}}
\!\!\!\! & =& \!\!\!\! 
\left[ A_d -  \mu_{\scriptscriptstyle 0}^* \, \tan\!\beta \right]\,m_{d}\;
+ \frac{\sqrt{2}\, M_{\!\scriptscriptstyle W} \cos\!\beta}
{g_{\scriptscriptstyle 2} } \,
\delta\! A^{\!{\scriptscriptstyle D}}_{\scriptscriptstyle 11}
\nonumber \\ \!\!\!\! &&
- \frac{\sqrt{2}\, M_{\!\scriptscriptstyle W} \sin\!\beta}
{g_{\scriptscriptstyle 2} } \,
(\, \mu_i^*\lambda^{\!\prime}_{i\scriptscriptstyle 11}\, ) \; .
\eeqa \normalsize
Note that the $\mu_i^*\lambda^{\!\prime}_{i\scriptscriptstyle 1\!1}$ term
does contain nontrivial CP violating phases and gives RPV contribution to
$d$ quark EDM.  Including inter-family mixings would
complicate the mass eigenstate analysis but not modify the EDM result in
any substantial way. We want to point out, without going into the details, 
that the analog of the type contribution to electron EDM, through a neutral 
(color-singlet) gaugino with a $\mu_i^* \, \lambda_{i\scriptscriptstyle 1\!1}$ 
slepton mixing, is obvious.

If one naively imposes the constraint for this RPV contribution itself not 
to exceed the experimental bound on neutron EDM, one gets roughly
$\mbox{Im}(\mu_i^*\lambda^{\!\prime}_{i\scriptscriptstyle 1\!1}) 
\lsim 10^{-6}\,\mbox{GeV}$, a constraint that is interesting even
in comparison to the bounds on the corresponding parameters obtainable
from asking no neutrino masses to exceed the super-Kamiokande (super-K)
atmospheric oscillation scale\cite{as4}. In fact, the most stringently interpreted
bounds on the individual parameters involved are given by
\[
\lambda^{\!\prime}_{\scriptscriptstyle 31\!1}\lsim \;0.05\sim 0.1
\]
and 
\[
\mu_i \; \cos\!\zb \lsim 10^{-4}\,\mbox{GeV} \; ,
\]
while the $\mu_i$ bounds admitting a heavier neutrinos are much weaker, 
as shown in Table~1.

\subsection{Fermion Mixings and EDM}
Once we see the above discussed EDM contribution through $LR$-scalar mixing,
it is quite natural to expect the same kind of RPV parameter combinations
could contribute in other diagrams. In fact,
there are other 1-loop contributions. In the case of MSSM, the chargino
contribution is known to be competitive or even dominates over the gluino
one in some regions of the parameter space\cite{KiOs}. The major part of the
chargino contribution comes from a diagram with a gauge and a Yukawa coupling
for the loop vertices, with pure $L$-squark running in the loop. Here we
give the corresponding formula generalized to the case of SUSY without 
R-parity. This is given by\cite{as6}
\small \beqa 
\left({d_{\scriptscriptstyle f} \over e} \right)_{\!\!\chi^{\!\!\mbox{ -}}} 
\!\!\!\!\!\!\!\! && \!\!\!\! = \; -
{\alpha_{\mbox{\tiny em}} \over 4 \pi \,\sin\!^2\theta_{\!\scriptscriptstyle W}} \; 
\sum_{\scriptscriptstyle \tilde{f}'\mp} 
\sum_{n=1}^{5} \,\mbox{Im}({\cal C}_{\!fn\mp}) \;
{{M}_{\!\scriptscriptstyle \chi^{\mbox{-}}_n} \over 
M_{\!\scriptscriptstyle \tilde{f}'\mp}^2} \;
\nonumber \\   &&  \!\!\!\!\!\!\!\! \cdot
\left[ {\cal Q}_{\!\tilde{f}'} 
B\!\left({{M}_{\!\scriptscriptstyle \chi^{\mbox{-}}_{n}}^2 \over 
M_{\!\scriptscriptstyle \tilde{f}'\mp}^2} \right) 
+ ( {\cal Q}_{\!{f}} - {\cal Q}_{\!\tilde{f}'} ) 
A\!\left({{M}_{\!\scriptscriptstyle \chi^{\mbox{-}}_{n}}^2 \over 
M_{\!\scriptscriptstyle \tilde{f}'\mp}^2} \right) 
\right]  ,
\nonumber \\   &&  \!\!\!\!\!\!\!\!\label{edmco}
\eeqa \normalsize
for $f$ being $u$ ($d$) quark and $f'$ being $d$ ($u$), where
\beqa
{\cal C}_{un\mp} \!\! &=&  \!\!
{y_{\!\scriptscriptstyle u} \over g_{\scriptscriptstyle 2} } \,\, 
\mbox{\boldmath $V$}^{\!*}_{\!\!2n} \, {\cal D}_{d1\mp} \;\;\;\cdot
\Big(  \mbox{\boldmath $U$}_{\!1n} \,{\cal D}^{*}_{d1\mp} 
 \nonumber \\   &&  \!\! \!\!
\left. -
{y_{\!\scriptscriptstyle d} \over g_{\scriptscriptstyle 2} }\,\, 
\mbox{\boldmath $U$}_{\!2n}\,  {\cal D}^{*}_{d2\mp}
- {\lambda^{\!\prime}_{\scriptscriptstyle i11} \over g_{\scriptscriptstyle 2} }\,\, 
\mbox{\boldmath $U$}_{\!(i+2)n}\,  {\cal D}^{*}_{d2\mp} \right) \; ,
\nonumber \\
{\cal C}_{dn\mp} \!\! &=& \!\!
\left( {y_{\!\scriptscriptstyle d} \over g_{\scriptscriptstyle 2} }\,\, 
\mbox{\boldmath $U$}_{\!2n} 
+ {\lambda^{\!\prime}_{\scriptscriptstyle i11} \over g_{\scriptscriptstyle 2} }\,\, 
\mbox{\boldmath $U$}_{\!(i+2)n} \right)\! {\cal D}_{u1\mp} \;
\nonumber \\   && \;\; \cdot
\left( \mbox{\boldmath $V$}^{\!*}_{\!\!1n} \,{\cal D}^{*}_{u1\mp} -
{y_{\!\scriptscriptstyle u} \over g_{\scriptscriptstyle 2} } \,
\mbox{\boldmath $V$}^{\!*}_{\!\!2n} \, {\cal D}^{*}_{u2\mp} \right) 
\; ,
\nonumber \\
&&  \mbox{\small(only repeated index $i$ is to be summed)} 
 \nonumber \\   && 
\label{Cnmp}
\eeqa
and the loop integral function $B(x)$ and $A(x)$ given by
\beqa \label{Bx}
B(x) \!\! &=& \!\!
{1 \over 2\,(x-1)^2} \left[1 + x + {2\,x \ln x \over (1-x) } \right] \; ,
\nonumber \\
A(x) \!\! &=& \!\!
 {1 \over 2 \, (1-x)^2} \left(3 - x + {2\ln x \over 1-x} \right) \; .
\eeqa
The fermion diagonalization matrices are defined as
$\mbox{\boldmath $V$}^\dag {\mathcal{M}_{\scriptscriptstyle C}} \,
\mbox{\boldmath $U$} = \mbox{diag} 
\{ {M}_{\!\scriptscriptstyle \chi^{\mbox{-}}_n} \}$; 
and {${\cal D}_{\!f}$} diagonalizes the $f$-squark.

The terms in ${\cal C}_{dn\mp}$ with only one factor of 
${1\over g_{\scriptscriptstyle 2}}$ and a 
$\lambda^{\!\prime}_{i\scriptscriptstyle 1\!1}$ gives the RPV analog of
the dominating MSSM chargino contribution. The term
is described by a diagram, which at first order requires a 
${l}_{\scriptscriptstyle L_i}^{\!\!\mbox{ -}}$--$\tilde{W}^{\scriptscriptstyle +}$ 
mass mixing. The latter vanishes, as shown in Eq.(\ref{mc}). From the full
formula above, it is easy to see that the exact mass
eigenstate result would deviate from zero only to the extent that the mass
dependence of the $B$ and $A$ functions spoils the GIM-like
cancellation in the sum. The resultant contribution, however, is shown by
our exact numerical calculation to be substantial.
What is most interesting here is that an analysis 
through perturbational approximations illustrates that the contribution
is proportional to, basically, the same combination of RPV parameters,
{\it i.e.} $\mu_i^* \, \lambda^{\!\prime}_{\scriptscriptstyle i11}$.
Readers are referred to Refs.\cite{as4,as6} for details.
Again, we want to point out that the analog of the type of chargino-like loop
contribution to electron EDM, through
 $\mu_i^* \, \lambda_{i\scriptscriptstyle 1\!1}$ is obvious.

\section{SLEPTON MASSES \hspace*{.1in} \\ \hspace*{.2in} AND PHENOMENOLOGY}
\subsection{The Charged Scalars}
Things in the slepton sector are more complicated.
We have eight charged scalar states, including an unphysical Goldstone mode,
The $8\times 8$ mass-squared matrix of the following $1+4+3$ form :
\beq \label{ME}
{\cal M}_{\!\scriptscriptstyle {E}}^2 =
\left( \begin{array}{ccc}
\widetilde{\cal M}_{\!\scriptscriptstyle H\!u}^2 &
\widetilde{\cal M}_{\!\scriptscriptstyle LH}^{2\dag}  & 
\widetilde{\cal M}_{\!\scriptscriptstyle RH}^{2\dag}
\\
\widetilde{\cal M}_{\!\scriptscriptstyle LH}^2 & 
\widetilde{\cal M}_{\!\scriptscriptstyle LL}^{2} & 
\widetilde{\cal M}_{\!\scriptscriptstyle RL}^{2\dag} 
\\
\widetilde{\cal M}_{\!\scriptscriptstyle RH}^2 &
\widetilde{\cal M}_{\!\scriptscriptstyle RL}^{2} & 
\widetilde{\cal M}_{\!\scriptscriptstyle RR}^2  
\end{array} \right) \; ;
\eeq
where
\small
\beqa
\widetilde{\cal M}_{\!\scriptscriptstyle H\!u}^2 
\!\!\!\! &=& \!\!\!\!
\tilde{m}_{\!\scriptscriptstyle H_{\!\scriptscriptstyle u}}^2
+ \mu_{\!\scriptscriptstyle \za}^* \mu_{\scriptscriptstyle \za}
+ M_{\!\scriptscriptstyle Z}^2\, \cos\!2 \beta 
\left[ \,\frac{1}{2} - \sin\!^2\theta_{\!\scriptscriptstyle W}\right]
\nonumber \\
\!\!\!\! && \!\!\!\! +
 M_{\!\scriptscriptstyle Z}^2\,  \sin\!^2 \beta \;
[1 - \sin\!^2 \theta_{\!\scriptscriptstyle W}]
\; ,
\nonumber \\
\widetilde{\cal M}_{\!\scriptscriptstyle LL}^2
\!\!\!\! &=& \!\!\!\! 
\tilde{m}_{\!\scriptscriptstyle {L}}^2 +
m_{\!\scriptscriptstyle L}^\dag m_{\!\scriptscriptstyle L}
+ (\mu_{\!\scriptscriptstyle \za}^* \mu_{\scriptscriptstyle \zb})
\nonumber \\
\!\!\!\! && \!\!\!\! + \;
M_{\!\scriptscriptstyle Z}^2\, \cos\!2 \beta 
\left[ -\frac{1}{2} +  \sin\!^2 \theta_{\!\scriptscriptstyle W}\right] 
\nonumber \\
\!\!\!\! && \!\!\!\!  \;\; + \;
 \left( \!\! \begin{array}{cc}
 M_{\!\scriptscriptstyle Z}^2\,  \cos\!^2 \beta \;
[1 - \sin\!^2 \theta_{\!\scriptscriptstyle W}] 
\!\!&\!\! \quad 0_{\scriptscriptstyle 1 \times 3} \quad \\
0_{\scriptscriptstyle 3 \times 1} \!\!&\!\! 0_{\scriptscriptstyle 3 \times 3}  
 \end{array} \!\!\!\! \!\!
\right) \; ,
\nonumber \\
\widetilde{\cal M}_{\!\scriptscriptstyle RR}^2 
\!\!\!\! &=& \!\!\!\!
\tilde{m}_{\!\scriptscriptstyle {E}}^2 +
m_{\!\scriptscriptstyle E} m_{\!\scriptscriptstyle E}^\dag
+ M_{\!\scriptscriptstyle Z}^2\, \cos\!2 \beta 
\left[  - \sin\!^2 \theta_{\!\scriptscriptstyle W}\right] \; , 
\nonumber \\
\label{ELH}
\widetilde{\cal M}_{\!\scriptscriptstyle LH}^2
\!\!\!\! &=& \!\!\!\!
(B_{\za}^*)  
+ \left( \begin{array}{c} 
{1 \over 2} \,
M_{\!\scriptscriptstyle Z}^2\,  \sin\!2 \beta \,
[1 - \sin\!^2 \theta_{\!\scriptscriptstyle W}]  \\
0_{\scriptscriptstyle 3 \times 1} 
\end{array} \right)\; ,
\nonumber \\
\label{ERH}
\widetilde{\cal M}_{\!\scriptscriptstyle RH}^2
\!\!\!\! &=& \!\!\!\!
  -\,(\, \mu_i^*\lambda_{i{\scriptscriptstyle 0}k}\, ) \; 
\frac{v_{\scriptscriptstyle 0}}{\sqrt{2}} 
\nonumber \\ 
\!\!\!\! &=& \!\!\!\!
(\, \mu_k^* \, m_k \, ) \hspace*{.8in} 
\mbox{\footnotesize (no sum over $k$)}  \; ,
\nonumber \\ 
\label{ERL}
(\widetilde{\cal M}_{\!\scriptscriptstyle RL}^{2})^{\scriptscriptstyle T} 
\!\!\!\! &=& \!\!\!\!
 \left(\begin{array}{c} 
0  \\   A^{\!{\scriptscriptstyle E}} 
\end{array}\right)
 \frac{v_{\scriptscriptstyle 0}}{\sqrt{2}}
-\,(\, \mu_{\scriptscriptstyle \za}^*
\lambda_{{\scriptscriptstyle \za\zb}k}\, ) \, 
\frac{v_{\scriptscriptstyle u}}{\sqrt{2}} \; ,
\nonumber \\
\!\!\!\! &=& \!\!\!\!
[A_e - \mu_{\scriptscriptstyle 0}^* \, \tan\!\beta ] 
\left(\begin{array}{c}  
0  \\   m_{\!{\scriptscriptstyle E}} 
\end{array}\right) \,
\nonumber \\
\!\!\!\! && 
+ \; \frac{\sqrt{2}\, M_{\!\scriptscriptstyle W} \cos\!\beta}
{g_{\scriptscriptstyle 2} } \,
\left(\begin{array}{c} 
0  \\ \delta\! A^{\!{\scriptscriptstyle E}}
\end{array}\right)
\nonumber \\
\!\!\!\! && \;\;\;\;
-  \; \left(\begin{array}{c}  
- \mu_{k}^* \, m_k\, \tan\!\beta \\ 
\frac{\sqrt{2}\, M_{\!\scriptscriptstyle W} \sin\!\beta}
{g_{\scriptscriptstyle 2} } \,(\, \mu_i^*\lambda_{ijk}\, ) 
\end{array}\right) \; .
\label{ERL}
\eeqa \normalsize
Notations and results here are similar to the squark case above, with some 
difference. We have $A_e$ and $\delta\! A^{\!{\scriptscriptstyle E}}$,
or the extended matrices {\tiny $\left(\begin{array}{c} 
0  \\   \star
\end{array}\right)$} incorporating them, denote the splitting of the $A$-term,
with proportionality defined with respect to 
$m_{\!\scriptscriptstyle E}$; $m_{\!\scriptscriptstyle L}=
\mbox{diag}\{0,m_{\!\scriptscriptstyle E}\}= \mbox{diag}\{0,m_{\!\scriptscriptstyle 1},
m_{\!\scriptscriptstyle 2},m_{\!\scriptscriptstyle 3}\}$. Recall that the $m_i$'s
are approximately the charged lepton masses. 

A $4\times 3$ matrix $(\, \mu_i^*\lambda_{i{\scriptscriptstyle \zb}k}\, )$
gives the RPV contributions to 
$(\widetilde{\cal M}_{\!\scriptscriptstyle RL}^{2})^{\scriptscriptstyle T}$
which is the $LR$-mixing part. 
In the above expression, we separate explicitly the first row of the former,
which corresponds to mass-squared terms of the type
$\tilde{l}^{\scriptscriptstyle +} h_{\scriptscriptstyle d}^{\!\!\mbox{ -}}$ 
type ($h_{\scriptscriptstyle d}^{\!\!\mbox{ -}} \equiv 
\tilde{l}_{\scriptscriptstyle 0}^{\!\!\mbox{ -}}$). This is the piece that
gives rise to the Zee neutrino mass diagram\cite{zee} within the present
SUSY framework, in which the $R$-handed sleptons play the role of the 
Zee scalar\cite{as1}. The remaining $3\times 3$ 
part given by  $(\, \mu_i^*\lambda_{ijk}\, )$ is the exact analog of the
squark mixings discussed above. We have already pointed out the contributions
to electron EDM from the RPV parameter combination
$\mu_i^* \, \lambda_{i\scriptscriptstyle 1\!1}$ in analog to that of the $d$
quark. Between the talk and the preparation of this proceedings submission,
we have also published a comprehensive study of the $\mu \to e\,\gamma$
process, in which the $\mu_i^* \, \lambda_{i\scriptscriptstyle 21}$ and
$\mu_i^* \, \lambda_{i\scriptscriptstyle 1\!2}$ combinations play an
important role\cite{as7}. 

Unlike the squarks,, however, we have also the
$\mu_i^* \,  \mu_j$ flavor changing $LL$-mixing in
$\widetilde{\cal M}_{\!\scriptscriptstyle LL}^{2}$. Among other things,
this also contributes to  $\mu \to e\,\gamma$\cite{as7}. The nonzero  
$\widetilde{\cal M}_{\!\scriptscriptstyle RH}^2$ and the $B_i^*$'s in 
$\widetilde{\cal M}_{\!\scriptscriptstyle LH}^2$ are also RPV contributions.
The former is a $\tilde{l}^{\scriptscriptstyle +} 
(h_{\scriptscriptstyle u}^{\scriptscriptstyle +})^{\dag} $
type, while the latter a $\tilde{l}^{\!\!\mbox{ -}} 
h_{\scriptscriptstyle u}^{\scriptscriptstyle +} $  term.
Note that the parts with the 
$[1 - \sin\!^2 \theta_{\!\scriptscriptstyle W}]$ factor are singled out 
as they are extra contributions to the masses of the charged-Higgses
({\it i.e.} $l_{\scriptscriptstyle 0}^{\!\!\mbox{ -}}
\equiv h_{\!\scriptscriptstyle d}^{\!\!\mbox{ -}}$ and
$h_{\!\scriptscriptstyle u}^{\!\scriptscriptstyle +}$).
\begin{figure}[b]
\includegraphics{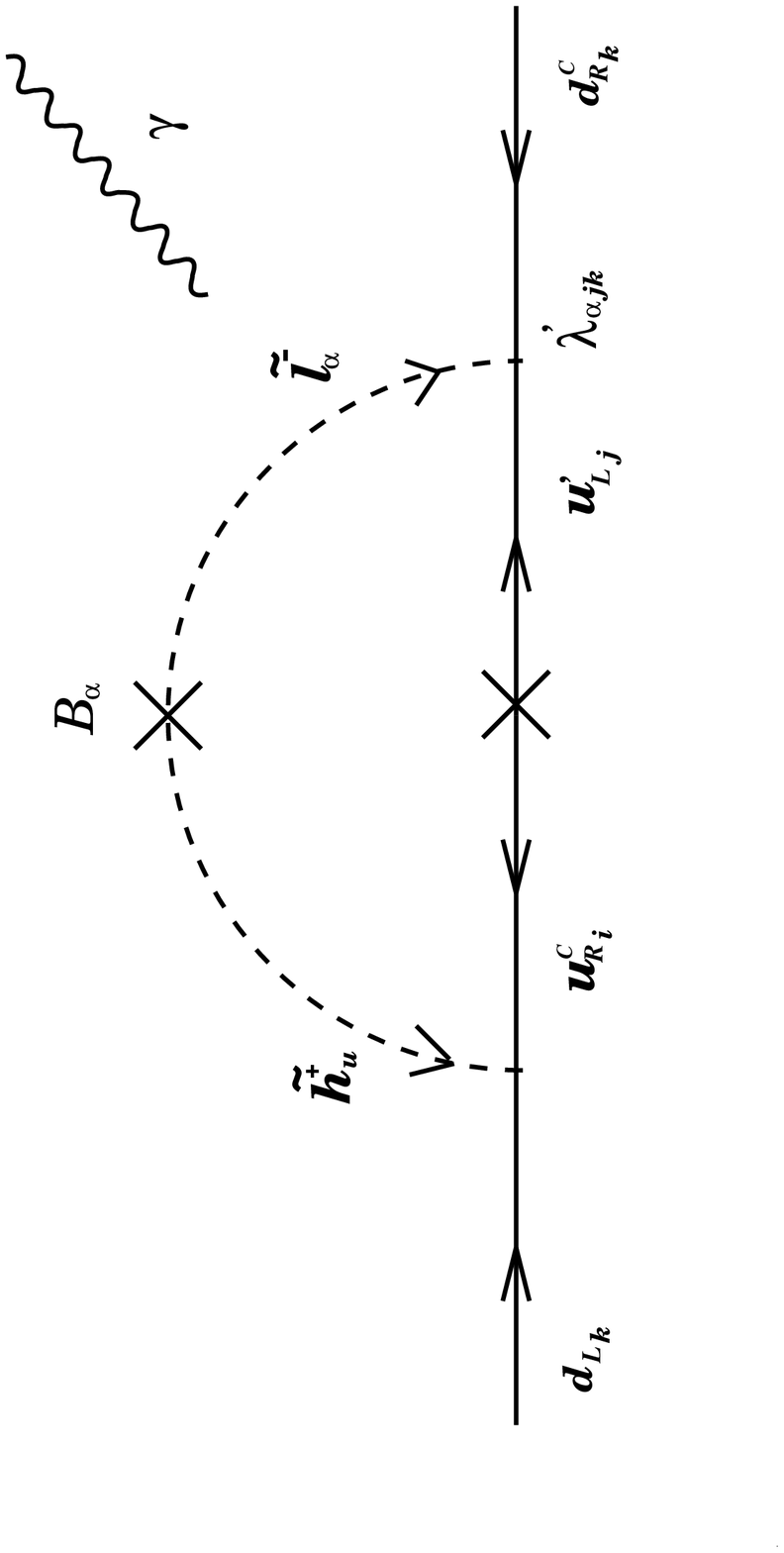}
\vspace*{1.2in}
\caption{\small Quark-scalar loop contribution to $d$ quark EDM.}
\end{figure}
The latter is the result of the quartic terms in the scalar potential and
the fact that the Higgs doublets bear VEV's. Such scalar mixings also play 
a role in contributing to EDM's. For example, that is a top loop contribution
to $d$ quark EDM of the form depicted in Fig.~1\cite{cch1,as6}.

\subsection{The Neutral Scalars}
The neutral scalar mass terms, in terms of the
$(1+4)$ complex scalar fields,  $\phi_n$'s, can be written in two parts
--- a simple $({\cal M}_{\!\scriptscriptstyle {\phi}{\phi}\dag}^2)_{mn} \,
\phi_m^\dag \phi_n$ part, and a Majorana-like part in the form 
${1\over 2} \,  ({\cal M}_{\!\scriptscriptstyle {\phi\phi}}^2)_{mn} \,
\phi_m \phi_n + \mbox{h.c.}$. As the neutral scalars are originated
from chiral doublet superfields, the existence of the Majorana-like
part is a direct consequence of the electroweak symmetry
breaking VEV's, hence restricted to the scalars playing the Higgs
role only. They come from the quartic terms of the Higgs fields in
the scalar potential. We have explicitly
\footnotesize \beqa \label{Mpp}
{\cal M}_{\!\scriptscriptstyle {\phi\phi}}^2 
\!\!\!\! &=& \!\!\!\!
{1\over 2} \, M_{\!\scriptscriptstyle Z}^2\,
\left( \begin{array}{ccc}
 \sin\!^2\! \beta  \!\! & \!\!  - \cos\!\beta \, \sin\! \beta
\!\! & \!\! \quad 0_{\scriptscriptstyle 1 \times 3} \\
 - \cos\!\beta \, \sin\! \beta \!\! & \!\! \cos\!^2\! \beta 
\!\! & \!\! \quad 0_{\scriptscriptstyle 1 \times 3} \\
0_{\scriptscriptstyle 3 \times 1} \!\! & \!\! 0_{\scriptscriptstyle 3 \times 1} 
\!\! & \!\! \quad 0_{\scriptscriptstyle 3 \times 3} 
\end{array} \right) ;
\nonumber \\ &&
\eeqa
{\normalsize and}
\small \beqa 
{\cal M}_{\!\scriptscriptstyle {\phi}{\phi}\dag}^2 
\!\!\!\! &=& \!\!\!\! 
 \left( \!\! \begin{array}{cc}
\tilde{m}_{\!\scriptscriptstyle H_{\!\scriptscriptstyle u}}^2
+ \mu_{\!\scriptscriptstyle \za}^* \mu_{\scriptscriptstyle \za}
  -\frac{1}{2} z
& - (B_\za) \\
- (B_\za^*) &
\tilde{m}_{\!\scriptscriptstyle {L}}^2 
+ (\mu_{\!\scriptscriptstyle \za}^* \mu_{\scriptscriptstyle \zb})
+ \frac{1}{2} z
\end{array} \!\! \right) 
\nonumber \\
\!\!\!\! &&  \;\; + \;
{\cal M}_{\!\scriptscriptstyle {\phi\phi}}^2 \; ,
\label{Mp}
\eeqa \normalsize
with
\beq
z = M_{\!\scriptscriptstyle Z}^2\, \cos\!2 \beta \; .
\nonumber
\eeq	
Note that ${\cal M}_{\!\scriptscriptstyle {\phi\phi}}^2$ here is 
real (see the appendix), while 
${\cal M}_{\!\scriptscriptstyle {\phi}{\phi}\dag}^2$ does have complex entries.
The full $10\times 10$ (real and symmetric) mass-squared matrix for 
the real scalars is then given by
\beq \label{MSN}
{\cal M}_{\!\scriptscriptstyle S}^2 =
\left( \begin{array}{cc}
{\cal M}_{\!\scriptscriptstyle SS}^2 &
{\cal M}_{\!\scriptscriptstyle SP}^2 \\
({\cal M}_{\!\scriptscriptstyle SP}^{2})^{\!\scriptscriptstyle T} &
{\cal M}_{\!\scriptscriptstyle PP}^2
\end{array} \right) \; ,
\eeq
where the scalar, pseudo-scalar, and mixing parts are
\beqa
{\cal M}_{\!\scriptscriptstyle SS}^2 &=&
\mbox{Re}({\cal M}_{\!\scriptscriptstyle {\phi}{\phi}\dag}^2)
+ {\cal M}_{\!\scriptscriptstyle {\phi\phi}}^2 \; ,
\nonumber \\
{\cal M}_{\!\scriptscriptstyle PP}^2 &=&
\mbox{Re}({\cal M}_{\!\scriptscriptstyle {\phi}{\phi}\dag}^2)
- {\cal M}_{\!\scriptscriptstyle {\phi\phi}}^2 \; ,
\nonumber \\
{\cal M}_{\!\scriptscriptstyle SP}^2 &=& -
\mbox{Im}({\cal M}_{\!\scriptscriptstyle {\phi}{\phi}\dag}^2) \; ,
\label{lastsc}
\eeqa
respectively. If $\mbox{Im}({\cal M}_{\!\scriptscriptstyle {\phi}{\phi}\dag}^2)$
vanishes, the scalars and pseudo-scalars decouple from one another and 
the unphysical Goldstone mode would be found among the latter. 

The most interesting part of the neutral scalar masses involves the
RPV parameters $B_i$'s and the corresponding mixing part in 
$\tilde{m}_{\!\scriptscriptstyle {L}}^2$. These parameters are not all 
independent, as discussed below in the next subsection. The $B_i$'s,
for example, lead to seesaw type Majorana-like mass terms for the 
``sneutrinos"\cite{Maj}, as depicted in Fig.~2.
\begin{figure}[t]
\includegraphics{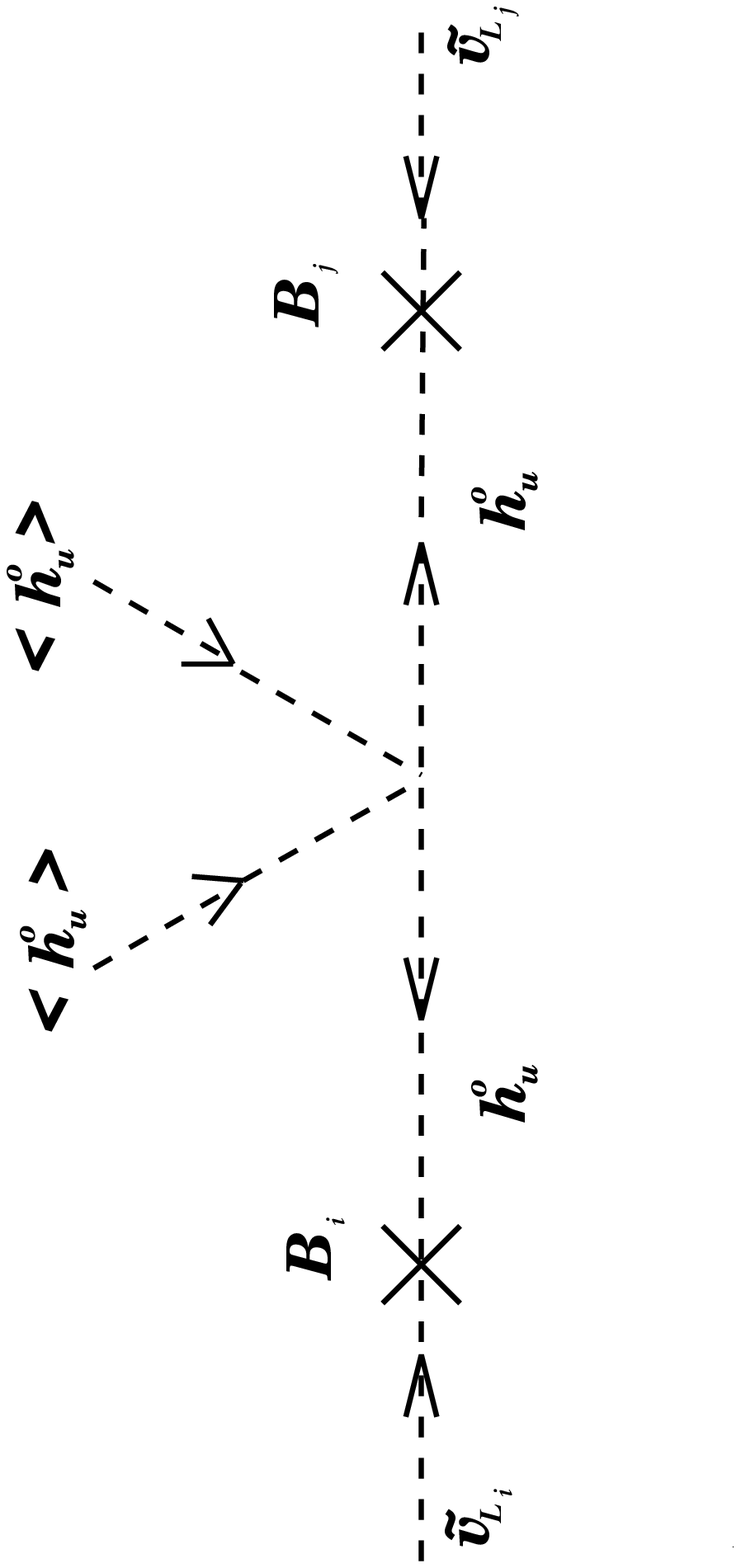}
\vspace*{.8in}
\caption{\small Seesaw origin of R-parity violating Majorana-like scalar masses.}
\end{figure}
\begin{figure}[htb]
\includegraphics{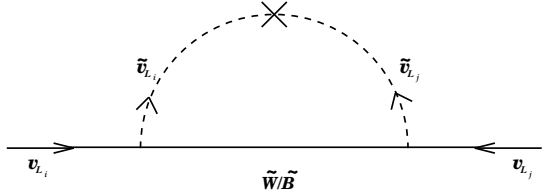}
\vspace*{.8in}
\caption{\small Gauge loop contribution to neutrino masses.}
\end{figure}
One of the interesting consequence here is a gauge loop contribution to
neutrino masses\cite{GH}, depicted in Fig.~3.

\subsection{A Look at the Scalar Potential}
We would like to emphasize that the above scalar mass results are complete 
--- all RPV contributions, SUSY breaking or otherwise, are included. The 
simplicity of the result is a consequence of the SVP.
Explicitly, there are no RPV $A$-term contributions due to the vanishing
of VEV's $v_i\equiv \sqrt{2}\langle\hat{L}_i\rangle$. The Higgs-slepton
results given as in Eqs.(\ref{ME}) and (\ref{MSN}) contain redundancy of 
parameters and hide the unphysical Goldstone states. The latter can be easily
identified, after implementation of the tadpole equations. The equations also
identify important relations among dependent parameters. 

In terms of the five, plausibly electroweak symmetry breaking, neutral 
scalars fields $\phi_n$, the generic (tree-level) scalar potential, as 
constrained by SUSY, can be written as :
\beqa
V_s \!\!\!\! &=& \!\!\!\! Y_{n} \left|\phi_n\right|^4 
+ X_{mn}  \left|\phi_m\right|^2 \left|\phi_n\right|^2 
+  \hat{m}^2_{n} \left|\phi_n\right|^2 
\nonumber \\
 \!\!\!\! && - ( \hat{m}^2_{\scriptscriptstyle m\!n} e^{i\theta\!_{m\!n}}
 \phi_m^{\dag} \phi_n + \mbox{h.c.}) \qquad \quad (m < n) \; .
\nonumber \\ &&
\eeqa
Here, we count the $\phi_n$'s from $-1$ to $3$ and identify a $\phi_\za$
(recall $\za=0$ to $3$) as 
$\tilde{l}_{\!\scriptscriptstyle \za}^{\scriptscriptstyle 0}$ 
and $\phi_{\scriptscriptstyle \mbox{-}\!1}$ as 
${h}_{\!\scriptscriptstyle u}^{\!\scriptscriptstyle 0}$.
Parameters in the above expression for $V_s$ (all real) are then given by
\beqa
 \hat{m}^2_{\za} \!\!\!\! &=& \!\!\!\! \tilde{m}^2_{\!{\scriptscriptstyle L\!_{\za\za}}}
 + \left|\mu_\za\right|^2 \; , \nonumber \\
 \hat{m}^2_{\scriptscriptstyle \mbox{-}\!1} \!\!\!\! &=& \!\!\!\! 
 \tilde{m}_{\!\scriptscriptstyle H\!_u}^2  + \mu_\za^{*} \mu_\za \; ,
\nonumber \\
\hat{m}^2_{\!\za\!\zb}\, e^{i\theta\!_{\za\!\zb}} \!\!\!\! &=& \!\!\!\! 
- \tilde{m}^2_{\!{\scriptscriptstyle L}\!_{ \za\!\zb}}
-  \mu_\za^{*} \mu_\zb \quad\quad  \mbox{(no sum)}\; ,	\nonumber \\
\hat{m}^2_{\!{\scriptscriptstyle \mbox{-}\!1\!\za}}\, 
e^{i\theta\!_{\scriptscriptstyle \mbox{-}\!1\!\za}} \!\!\!\! &=& \!\!\!\! B_\za 
\qquad\qquad\qquad\quad  \mbox{(no sum)}\; , \nonumber \\
 Y_{n} \!\!\!\! &=& \!\!\!\! 	\frac{1}{8}(g^2_{\scriptscriptstyle 1} 
 + g_{\scriptscriptstyle 2}^2)\; ,	\nonumber \\
 X_{{\scriptscriptstyle \mbox{-}\!1\!\za}}  \!\!\!\! &=& \!\!\!\! - 
\frac{1}{4}(g_{\scriptscriptstyle 1}^2 
 + g_{\scriptscriptstyle 2}^2) = -
 X_{{\scriptscriptstyle \za\!\zb}} \;.
\eeqa
Under the SVP, we write the VEV's as follows :
\beqa 
v_{\scriptscriptstyle \mbox{-}\!1}\,  (\equiv \sqrt{2}\,
\lla \phi_{\scriptscriptstyle \mbox{-}\!1} \rra)
\!\!\!\! & =& \!\!\!\!
v_{\scriptscriptstyle u} \; , \nonumber \\
v_{\scriptscriptstyle 0} \, (\equiv \sqrt{2}\,
\lla \phi_{\scriptscriptstyle 0} \rra)
\!\!\!\! & =& \!\!\!\!
v_{\scriptscriptstyle d}\, e^{i\theta\!_v} \; , \nonumber \\
v_{\scriptscriptstyle i} \, (\equiv \sqrt{2}\,
\lla \phi_i \rra)
\!\!\!\! & =& \!\!\!\! 0 \; ,
\eeqa 
where we have put in a complex phase in the VEV $v_{\scriptscriptstyle 0}$,
for generality. 

The equations from the vanishing derivatives of $V_s$ along
$\phi_{\scriptscriptstyle \mbox{-}\!1}$ and $\phi_{\scriptscriptstyle 0}$
give
\beqa
{ \left[ \frac{1}{8}(g_{\scriptscriptstyle 1}^2  + g_{\scriptscriptstyle 2}^2) 
(v_{\scriptscriptstyle u}^2 -v_{\scriptscriptstyle d}^2) + 
\hat{m}^2_{\scriptscriptstyle \mbox{-}\!1} \right]} 
\, v_{\scriptscriptstyle u}
\!\!\!\! &=& \!\!\!\!
B_0 \, v_{\scriptscriptstyle d} \, e^{i\theta\!_v} \; ,
\nonumber \\
{ \left[ \frac{1}{8}(g_{\scriptscriptstyle 1}^2  + g_{\scriptscriptstyle 2}^2) 
(v_{\scriptscriptstyle d}^2 -v_{\scriptscriptstyle u}^2) + 
\hat{m}^2_{\scriptscriptstyle 0} \right] }
\, v_{\scriptscriptstyle d}
\!\!\!\! &=& \!\!\!\!
B_0 \, v_{\scriptscriptstyle u} \, e^{i\theta\!_v} \; .
\eeqa
Hence, $B_0 \, e^{i\theta\!_v}$ is real. In fact, the part of $V_s$ that
is relevant to obtaining the tadpole equations is no different from
that of MSSM apart from the fact that 
$\tilde{m}_{\!\scriptscriptstyle H\!_u}^2$ and 
$\tilde{m}_{\!\scriptscriptstyle H\!_d}^2$ of the latter are replaced by
$\hat{m}^2_{\scriptscriptstyle \mbox{-}\!1}$ and
$\hat{m}^2_{\scriptscriptstyle 0}$ respectively. As in MSSM, the $B_0$
parameter can be taken as real. The conclusion here
is therefore that $\theta\!_v$ vanishes, or all VEV's are real, despite
the existence of complex parameters in the scalar potential.
Results from the other tadpole equations, in a $\phi_i$ direction, are quite
simple. They can be written as complex equations of the form 
\begin{equation}
 \hat{m}^2\!\!_{{\scriptscriptstyle \mbox{-}\!1}i}\; 
e^{i\theta\!_{{\scriptscriptstyle \mbox{-}\!1}i}} \tan\!\beta
= -  e^{i\theta\!_v} \;
 \hat{m}^2\!\!_{{\scriptscriptstyle 0}i}\; 
e^{i\theta\!_{{\scriptscriptstyle 0}i}} \; ,
\end{equation}
which is equivalent to 
\begin{equation}
B_i \, \tan\!\beta 
=  \tilde{m}^2_{{\scriptscriptstyle L}_{\!{\scriptscriptstyle 0}i} }
+ \mu_{\scriptscriptstyle 0}^{*} \, \mu_i \; ,
\end{equation} 
where we have used $v_{\scriptscriptstyle u}=v\sin\!\beta$ and
$v_{\scriptscriptstyle d}=v\cos\!\beta$. Note that our $\tan\!\beta$ has 
the same physical meaning as that in the R-parity conserving case. For
instance, $\tan\!\beta$, together with the corresponding Yukawa coupling 
ratio, gives the mass ratio between the top and the bottom quark.

The three complex equations for the $B_i$'s reflect the redundance of 
parameters in a generic $\hat{L}_\za$ flavor basis. The equations also 
suggest that the $B_i$'s are expected to be suppressed, with respect to
the R-parity conserving $B_{\scriptscriptstyle 0}$, as the $\mu_i$'s are,
with respect to $\mu_{\scriptscriptstyle 0}$. They give consistence
relationships among the involved RPV parameters (under the SVP) that
should not be overlooked.

\section{CONCLUDING REMARKS}
The complete theory of SUSY without R-parity is just the generic supersymmetric
SM. Giving up R-parity is particularly well-motivated by the expectation
of Majorana neutrino masses. The model gives many interesting results, that would 
otherwise be missed in limited version of RPV models. It is important to have 
a consistent and efficient framework to deal with the various kinds of RPV 
parameters; a specific and optimal parametrization of the model is needed to match 
parameters unambiguously with experiemental data. Our formulation (SVP) reviewed
here provides such a parametrization. It simplifies all (tree-level) mass matrices
very substantially, hence giving a strong technical advantage to  
phenomenological studies of the model. We have summarized some of our interesting
new result here. Many other features of the model awaits careful analysis.

%% file: as12t.tex


\begin{table*}[h]
\setlength{\tabcolsep}{.8pc}
\newlength{\digitwidth} \settowidth{\digitwidth}{\rm 0}
\catcode`?=\active \def?{\kern\digitwidth}
\caption{Summary of Phenomenological Constraints from Leptonic
$Z^0$ and $W^\pm$ Couplings.}
\begin{center}  
 \small
  \begin{tabular*}{\textwidth}{||l|c|c||} 
\hline \hline 
       {\quad Quantity \quad} & 
            $\stackrel{{\mu}_{i}\hbox{ combo.}}{
             { \scriptstyle { \hbox{constrained}}}}$ &
       {\quad Experimental bounds \quad} \\[.05in] \hline
       & & \\[-.1in]
       \framebox{$Z^0$-coupling:} & & \\[.05in]
      $\bullet$ $Br$($\mu^- \to e^- e^+e^-$)
       &  $|\mu_{\scriptscriptstyle 1}\mu_{\scriptscriptstyle 2}|$ 
       &  $<1.0\times 10^{-12}$ \\[.05in]
        $\bullet$ $Br$($\tau^- \to e^- e^+e^-$)
       & $|\mu_{\scriptscriptstyle 1}\mu_{\scriptscriptstyle 3}|$ 
       & $<2.9\times 10^{-6}$\\[.05in]
        $\bullet$ $Br$($\tau^{-} \to \mu^{-} e^+ e^-$)
       & $|\mu_{\scriptscriptstyle 2}\mu_{\scriptscriptstyle 3}|$ 
       &  $<1.7\times 10^{-6}$\\[.05in]
       $\bullet$ $Br$($\tau^{-} \to \mu^{+} e^- e^-$)
       &  $|\mu_{\scriptscriptstyle 1}^2\mu_{\scriptscriptstyle 2}\mu_{\scriptscriptstyle 3}|$ 
       & $<1.5\times 10^{-6}$\\[.05in]
       $\bullet$ $Br$($\tau^{-} \to e^{-} \mu^+ \mu^-$)
       &  $|\mu_{\scriptscriptstyle 1}\mu_{\scriptscriptstyle 3}|$ 
       & $<1.8\times 10^{-6}$\\[.05in]
       $\bullet$ $Br$($\tau^{-} \to e^{+} \mu^- \mu^-$)
       &  $|\mu_{\scriptscriptstyle 1}\mu_{\scriptscriptstyle 2}^2\mu_{\scriptscriptstyle 3}|$ 
       & $<1.5\times 10^{-6}$\\[.05in]
       $\bullet$ $Br$($\tau^- \to \mu^- \mu^+ \mu^-$)
       &  $|\mu_{\scriptscriptstyle 2}\mu_{\scriptscriptstyle 3}|$ 
       & $<1.9\times 10^{-6}$\\[.05in]
       $\bullet$ $Br$($Z^0 \to e^{\pm} \mu^{\mp}$)
       &  $|\mu_{\scriptscriptstyle 1}\mu_{\scriptscriptstyle 2}|$ 
       &  $<1.7\times 10^{-6}$ \\[.05in]
       $\bullet$ $Br$($Z^0 \to e^{\pm} \tau^{\mp}$)
       &  $|\mu_{\scriptscriptstyle 1}\mu_{\scriptscriptstyle 3}|$ 
       &  $<9.8\times 10^{-6}$ \\[.05in]
       $\bullet$ $Br$($Z^0 \to \mu^{\pm} \tau^{\mp}$)
       &  $|\mu_{\scriptscriptstyle 2}\mu_{\scriptscriptstyle 3}|$ 
       &  $<1.2\times 10^{-5}$ \\[.05in]
       $\bullet$ $Br$($Z^0 \to \chi^{\pm} \ell^{\mp}$)
       &  {complicated} 
       &  $< 1.0\times 10^{-5}$ \\[.05in]
       $\bullet$ $Br$($Z^0 \to \chi^{\pm} \chi^{\mp}$)
       &  $\mu_{\scriptscriptstyle 5}$ 
       &  $< 1.0\times 10^{-5}$ \\[.05in]
       $\bullet$ $U_{br}^{e\mu}$ \hspace*{.2in} ($e$-$\mu$ universality)
      &  $\mu_{\scriptscriptstyle 1}^2-\mu_{\scriptscriptstyle 2}^2$ 
       &  $(0.596 \pm 4.37)\times 10^{-3}$ \\[.05in]
       $\bullet$ $U_{br}^{e\tau}$ \hspace*{.2in} ($e$-$\tau$ universality)
       &  $\mu_{\scriptscriptstyle 1}^2-\mu_{\scriptscriptstyle 3}^2$ 
       &  $(0.955 \pm 4.98)\times 10^{-3}$ \\[.05in]
       $\bullet$  $U_{br}^{\mu\tau}$ \hspace*{.2in} ($\mu$-$\tau$ universality)
       &  $\mu_{\scriptscriptstyle 2}^2-\mu_{\scriptscriptstyle 3}^2$ 
       &  $(1.55 \pm 5.60)\times 10^{-3}$ \\[.05in]
      $\bullet$ $\Delta{\cal A}_{e\mu}$ \hspace*{.15in} ($e$-$\mu$ L-R asymmetry)
      &  $(\mu_{\scriptscriptstyle 1}^2-\mu_{\scriptscriptstyle 2}^2$) + Rt. contribution
      &  $(0.346\pm 2.54)\times 10^{-2}$ \\[.05in]
       $\bullet$ $\Delta{\cal A}_{\tau e}$ \hspace*{.1in} ($\tau$-$e$ L-R asymmetry)
       &  ($\mu_{\scriptscriptstyle 3}^2-\mu_{\scriptscriptstyle 1}^2$) + Rt. contribution
       &  $0.0043\pm 0.104$ \\[.05in]
       $\bullet$ $\Delta{\cal A}_{\tau\mu}$ \hspace*{.1in} ($\tau$-$\mu$ L-R asymmetry)
       &  ($\mu_{\scriptscriptstyle 3}^2-\mu_{\scriptscriptstyle 2}^2$) + Rt. contribution
       &  $0.082\pm 0.25$ \\[.05in]
       $\bullet$ ${\Gamma}_{\!\scriptscriptstyle Z}$ \hspace*{.3in} (total $Z^0$-width)
       &  $\mu_{\scriptscriptstyle 5}$
       &  $2.4948\pm.0075 \, \hbox{GeV}$ \\[.05in]
       $\bullet$ ${\Gamma}_{\!\scriptscriptstyle Z}^{\scriptscriptstyle i\!n\!v}$\hspace*{.3in} (*)
       &  $\mu_{\scriptscriptstyle 5}$ 
       &  $500.1\pm5.4\, \hbox{MeV}$ \\[.05in]
        $\bullet$ $Br$($Z^0 \to \chi^{0}_i\chi^{0}_j,
                                               \chi^{0}_j\nu) ; \; j \ne 1$
       &  $\mu_{\scriptscriptstyle 5}$ 
       &  $< 1.0\times 10^{-5}$ \\[.05in]
      & & \\[-.1in]
       \framebox{$W^{\pm}$-coupling:}
        & & \\[.05in]
       $\bullet$        $\overline{\Gamma}^{\mu e}$ \hspace*{.2in} ($\mu \to e \nu \nu$)
       &  $m_{\nu_{\scriptscriptstyle 3}}\,/\,\mu_i$ ratio  &
        $0.983\pm0.111$ \\[.05in]
       $\bullet$        $\overline{\Gamma}^{\tau e}$ \hspace*{.2in} ($\tau \to e \nu \nu$)
       &  $m_{\nu_{\scriptscriptstyle 3}}\,/\,\mu_i$ ratio  &
        $0.979\pm0.111$ \\[.05in]
       $\bullet$        $\overline{\Gamma}^{\tau \mu}$ \hspace*{.2in} ($\tau \to \mu \nu \nu$)
       &  $m_{\nu_{\scriptscriptstyle 3}}\,/\,\mu_i$ ratio  &
        $0.954\pm0.108$ \\[.05in]
       $\bullet$        $R^{\pi e}_{\pi \mu}$ \hspace*{.2in} ($\pi$ decays)
       &  $m_{\nu_{\scriptscriptstyle 3}}\,/\,\frac{\mu_{\scriptscriptstyle 1}}{\mu_{\scriptscriptstyle 5}}$ and $\frac{\mu_{\scriptscriptstyle 2}}{\mu_{\scriptscriptstyle 5}}$   &
        $(1.230\pm0.012)\times 10^{-4}$ \\[.05in]
           $\bullet$        $R^{\tau e}_{\tau \mu}$ \hspace*{.2in} ($\tau$ decays)
       &  $m_{\nu_{\scriptscriptstyle 3}}\,/\,\mu_i$ ratio   &
        $1.0265\pm0.0222$ \\[.05in]
        $\bullet$        $R^{\mu e}_{\tau e}$ \hspace*{.2in} (decays to $e$'s)
       &  $m_{\nu_{\scriptscriptstyle 3}}\,/\,\mu_i$ ratio &
        $1.0038\pm0.0219$ \\[.05in]
       $\bullet$ $m_{\nu_{\scriptscriptstyle 3}}
|\tilde{B}^{\scriptscriptstyle L}_{e\nu_{\scriptscriptstyle 3}}|^2$ \hspace*{.2in} [$(\zb\zb)_{0\nu}$]
       &  $m_{\nu_{\scriptscriptstyle 3}}\,/\,\frac{\mu_{\scriptscriptstyle 1}}{\mu_{\scriptscriptstyle 5}}$  &
        $< 0.68\, \hbox{eV}$  (only for $m_{\nu_{\scriptscriptstyle 3}}\!<\!10\,\mbox{MeV}$) \\[.05in]
        $\bullet$ BEBC expt.& $m_{\nu_{\scriptscriptstyle 3}}\,/\,\frac{\mu_{\scriptscriptstyle 1}}{\mu_{\scriptscriptstyle 5}}$ and $\frac{\mu_{\scriptscriptstyle 2}}{\mu_{\scriptscriptstyle 5}}$
       & \\[.05in]
      & & \\[-.1in]
       \framebox{mass constraints:}
       & & \\[.05in]
        $\bullet$ $\nu_{\scriptscriptstyle 3}$ mass
       &  $\mu_{\scriptscriptstyle 3}$ 
       &  $<18.2\, \hbox{MeV}$ if $\nu_{\scriptscriptstyle 3} = \nu_{\tau}$ 
       \\[.05in]
      &  $\mu_{\scriptscriptstyle 5}$  
    &  $<149\, \hbox{MeV}$ if $\nu_{\scriptscriptstyle 3} \ne \nu_{\tau}$ \\[.05in]
  \hline \hline
\end{tabular*}
\end{center}
\end{table*}

\normalsize

%% file: as12n.bbl
\begin{thebibliography}{99}
\bibitem{AV}
D.V. Volkov and V.P. Akulov, Phys. Lett. {\bf 46B}, 109 (1973).
\bibitem{Fy2}
G. Farrar and P. Fayet,  Phys. Lett. {\bf 76B}, 575 (1978).
\bibitem{osc}
Neutrino masses are expected from oscillation analyses of experimental anomalies.
The subject is well documented. For a recent comprehensive oscillation analysis, see
M. Gonzalez-Garcia, M. Maltoni, C. Pena-Garay, and J. Valle,
Phys. Rev. {\bf D63}, {\it 033005} (2001).
\bibitem{pd}
L.E. Ib\'a\~nez and G.G. Ross,
Nucl. Phys. {\bf B368}, 3 (1992).
\bibitem{as8} 
O.C.W. Kong, {\it  IPAS-HEP-k008}, {\it manuscript in preparation}.
\bibitem{as9}
S.K. Kang and O.C.W. Kong, {IPAS-HEP-k009},
{\it manuscript in preparation}.
See also Refs.\cite{ok,as1,as5,GH,AL}.
\bibitem{ru10}
O.C.W. Kong, Phys. Atom. Nucl. {\bf 63}, 1083 (2000).
\bibitem{ru}
M. Bisset, O.C.W. Kong, C. Macesanu, and L.H. Orr,
Phys. Lett. {\bf B430}, 274 (1998); 
Phys. Rev. {\bf D62},  {\it 035001} (2000).
\bibitem{ok}
O.C.W. Kong, Mod. Phys. Lett. {\bf A14},  903 (1999).
\bibitem{as1}
K. Cheung and O.C.W. Kong,  
Phys. Rev. {\bf D61},  {\it 113012} (2000).
\bibitem{zee}
A. Zee,  Phys. Lett. {\bf 93B}, 389 (1980).
\bibitem{as5}
O.C.W. Kong, JHEP {\bf 0009}, {\it 037} (2000).
\bibitem{as4}
Y.-Y. Keum and O.C.W. Kong, Phys. Rev. Lett. {\bf 86}, 393 (2001).
\bibitem{cch1}
K. Choi, E.J. Chun, and K. Hwang, Phys. Rev. {\bf D63},  {\it 013002} (2000).
\bibitem{as6}
Y.-Y. Keum and O.C.W. Kong,  {\it IPAS-HEP-k006}, hep-ph/0101113.
\bibitem{bs}
O.C.W. Kong {\it el.al.}, {\it work in progress}.
\bibitem{as7}
K. Cheung and O.C.W. Kong,   {\it IPAS-HEP-k007}, hep-ph/0101347;
see also K. Choi, E.J. Chun, and K. Hwang, Phys. Lett. {\bf B488}, 145 (2000). 
\bibitem{KiOs}
Y. Kizukuri and N. Oshimo, Phys. Rev. {\bf D46}, 3025 (1992).
\bibitem{Maj}
Y. Grossman and H.E. Haber,   hep-ph/9906310; see also 
M. Hirsch, H.V. Klapdor-Kleingrothaus S.G. Kovalenko, Phys. Lett. {\bf B 398}, 311
(1997).
\bibitem{GH}
Y. Grossman and H.E. Haber, Phys. Rev. {\bf D59},  {\it 093008} (1999);
see also Refs.\cite{as5,AL}.
\bibitem{AL}
A. Abada and M. Losada,  hep-ph/9908352;
S. Davidson and M. Losada, JHEP {\bf 0005}, {\it 021} (2000).

\end{thebibliography}
